\documentclass[twocolumn,showpacs,amsmath,amssymb]{revtex4-1}

\usepackage{graphicx}
\usepackage{dcolumn}
\usepackage{bm}
\usepackage[abs]{overpic}
\def\up{\uparrow}
\def\down{\downarrow }
\def\Vec#1{\bm{#1}}

\newcommand{\veck}{\bm{k}}

\begin{document}


\title{
Non-fragile superconductivity with nodes in the superconducting topological insulator Cu$_{x}$Bi$_{2}$Se$_{3}$: Zeeman orbital field and non-magnetic impurities
}

\author{Yuki Nagai}
\affiliation{CCSE, Japan  Atomic Energy Agency, 178-4-4, Wakashiba, Kashiwa, Chiba, 277-0871, Japan}

\date{\today}
             
\begin{abstract}
We study the robustness against non-magnetic impurities in the topological superconductor with point nodes, focusing on an effective model of Cu$_{x}$Bi$_{2}$Se$_{3}$. 
We find that the topological superconductivity with point-nodes is not fragile against non-magnetic impurities, although the superconductivity with nodes in past studies is usually fragile. 
Exchanging the role of spin with the one of orbital, and vice versa, we find that in the ``dual'' space the topological superconductor with point-nodes is regarded as the intra-orbital spin-singlet $s$-wave one. 
From the viewpoint of the dual space, we deduce that the point-node state is not fragile against non-magnetic impurity, when the orbital imbalance in the normal states is small. 
Since the spin imbalance is induced by the Zeeman magnetic field, we shall name this key quantity for the impurity effects Zeeman ``orbital'' field. 
The numerical calculations support that the deduction is correct. 
If the Zeeman orbital field is small, the topological superconductivity is not fragile in dirty materials, even with nodes.  
Thus, the topological superconductors can not be simply regarded as one
of the {\it conventional} unconventional superconductors.

\end{abstract}

\maketitle

The discovery of topological
insulators
\cite{Bernevig15122006,PhysRevLett.105.266401,PhysRevB.76.045302,PhysRevLett.98.106803,PhysRevLett.95.146802,Konig02112007,PhysRevLett.105.146801,PhysRevB.75.121306,PhysRevB.81.041309,PhysRevLett.105.136802}
leads to a number of the studies 
about topological aspects in solid-state
physics\cite{RevModPhys.82.3045}. 
Topological superconductors\cite{PhysRevLett.105.097001} are of
particular interest, since the emergence of the superconducting order is
associated with the occurrence of a non-trivial topological 
invariant\cite{PhysRevLett.95.146802,PhysRevB.75.121306,PhysRevB.76.045302,PhysRevLett.98.106803}. 
In addition, they allow us to manipulate the Majorana fermion in materials and open an intriguing way of quantum
engineering.

The quest for the bulk topological superconductors is an exciting issue
in topological material science. 
The copper intercalated topological insulator
$\mbox{Cu}_{x}\mbox{Bi}_{2}\mbox{Se}_{3}$ shows superconductivity at
$T_{\rm c}\approx 3.8\,\mbox{K}$ and is a 
candidate for the bulk topological superconductors
\cite{PhysRevLett.104.057001,Wray2010,MKR_L11,PhysRevLett.105.097001}. 
Identifying the gap-function type is now in great demand. 
The point-contact
spectroscopy\cite{PhysRevLett.107.217001,PhysRevB.86.064517} showed the
zero-bias conductance peaks from the Majorana bound states at the
surface edges.
However, the scanning tunneling
spectroscopy\cite{PhysRevLett.110.117001} 
indicated a fully-gapped feature in the density of states (DOS); 
there is no in-gap state, and therefore the superconducting
state could be topologically trivial. 
In addition, the Knight-shift measurement\cite{Matano} showed the
presence of in-plane anisotropy. 
Hashimoto \textit{et al.}\cite{Hashimoto} pointed out that the
anisotropy is related to a character of a point-nodes gap function on $a$-$b$-plane, since the
electronic structure in the normal states is almost isotropic. 
The point-node gap function also induces the in-plane anisotropy of the
thermal conductivity\cite{NagaiThermal}. 
Fu~\cite{Fupoint} argued a different scenario for the in-plane 
anisotropy, using an odd-parity full gap
state and the normal-state Hamiltonian with a hexagonal warping
term of the spin-orbital coupling. 

The possibility of the nodal gap function in
$\mbox{Cu}_{x}\mbox{Bi}_{2}\mbox{Se}_{3}$ is very surprising and
curious. 
Typically, this superconducting compound is
considered to be dirty owing to the copper intercalated process. 
Indeed, the short mean-free path was reported
experimentally\cite{MKR_L11}. 
A large number of the studies about superconducting
alloys\cite{Kopnin:2001,Hirschfeld;Woelfe:1986,SchmittRink;Varma:1986,Hotta:1993,Preosti;Muzikar:1996}
indicate that the superconductivity of an unconventional state (e.g.,
$d$-wave and chiral $p$-wave) promptly diminishes via impurity
scattering, different from the robustness of an $s$-wave state against
non-magnetic impurities (Anderson's theorem\cite{Kopnin:2001}). 
In particular, a nodal order is very fragile against
non-magnetic impurities, since the low-energy excitations are produced
around the nodes in the momentum space. 
Therefore, many questions arise: 
How does one understand the existence of a nodal superconducting state
in the dirty materials? 
Is the topological superconductor with a point node really fragile against
non-magnetic impurities?

In this paper, we study the robustness of a point-node gap function
against non-magnetic impurities in an effective model for
$\mbox{Cu}_{x}\mbox{Bi}_{2}\mbox{Se}_{3}$. 
The model has the massive Dirac Hamiltonian in the normal part and the
on-site pair potential in the superconducting part. 
We propose a simple and intuitive way of understanding the impurity
effects of a point-node gap function.
We focus on the spin and orbital degrees of freedom in the Bogoliubov-de
Gennes (BdG) Hamiltonian. 
Exchanging the role of spin with the one of orbital, and vise versa, one
can obtain a ``dual'' space with respect to the original space. 

From the viewpoint of the dual space, we deduce that the point-node state is
\textit{not} fragile against non-magnetic impurity, when the orbital
imbalance in the normal states is small. 
Since the spin imbalance is induced by the Zeeman magnetic field, we
shall name this key quantity for the impurity effects Zeeman
``orbital'' field. 
We find that the Zeeman orbital field is connected with the mass
of the Dirac Hamiltonian. 
Thus, the role of the
Zeeman orbital field is qualitatively examined by measuring the amount
of relativistic effects\cite{NagaiPRBimp}. 
The numerical calculations of the DOS support that the deduction is correct. 
Using a self-consistent $T$-matrix approach for impurity scattering
with a unitary limit, we confirm that the in-gap states in the DOS are not
induced, when the Zeeman orbital field is small (i.e., the normal-state
stays at the relativistic regime). 
In addition, we show the importance of the Zeeman orbital field for the
impurity effects of the point-node state in terms of the violation of
Anderson's theorem, within the Born approximation. 
Thus, the topological superconductors can not be simply regarded as one
of the {\it conventional} unconventional superconductors.

The mean-field Hamiltonian for $\mbox{Cu}_{x}\mbox{Bi}_{2}\mbox{Se}_{3}$
is 
\mbox{
$H = \int d^{3}\Vec{k}\, \Vec{\psi}^{\dagger}(\Vec{k}) {\cal H}(\Vec{k})
\Vec{\psi}(\Vec{k})$}. 
The 8-component column vector $\Vec{\psi}(\Vec{k})$ is composed of the
electron annihilation ($c_{\Vec{k},\alpha}$) and creation operators
($c^{\dagger}_{\Vec{k},\alpha}$), where $\alpha$ is a collective
coordinate for orbital ($1,\,2$) and spin ($\uparrow,\,\downarrow$);   
$\Vec{\psi}(\Vec{k}) = (c_{\Vec{k},1,\up},
c_{\Vec{k},2,\up},c_{\Vec{k},1,\down},
c_{\Vec{k},2,\down},c^{\dagger}_{-\Vec{k},1,\up},
c^{\dagger}_{-\Vec{k},2,\up},
c_{-\Vec{k},1,\down}^{\dagger},c_{-\Vec{k},2,\down}^{\dagger})^{\rm T}$.  
The $8\times 8$ matrix ${\cal H}(\Vec{k})$ is the BdG Hamiltonian
matrix\cite{NagaiThermal,NagaiQuasi,Hao,Yamakage,Mizushima}, 
\begin{align}
{\cal H}(\Vec{k}) &= 
\left(\begin{array}{cc}
h_{0}(\Vec{k}) & \Delta_{\rm pair}(\Vec{k})  \\
\Delta_{\rm pair}^{\dagger}(\Vec{k})  & - h_{0}^{\ast}(-\Vec{k})
\end{array}\right) .
\end{align}
The normal-part effective Hamiltonian $h_{0}$ is described by the
massive Dirac Hamiltonian with the strong spin-orbital coupling and the
negative Wilson mass term\cite{NagaiQuasi}, 
\begin{align}
h_{0}(\Vec{k})
= 
\epsilon(\Vec{k}) s^{0} \otimes \sigma^{0} 
+ h_{z}(\Vec{k}) + h_{so}(\Vec{k}), 
\end{align}
with 
$h_{z}(\Vec{k}) 
= 
M(\Vec{k}) s^{0} \otimes  \sigma^{3}$ and  
$h_{so}(\Vec{k})  
= \sum_{i=1}^{3} P_{i}(\Vec{k})s^{i} \otimes \sigma^{1}$,  
where $\sigma^{i}$ ($s^{i}$) are the $2\times 2$ Pauli matrices in the
orbital (spin) space. 
The identity matrix in each space is labeled by the superscript $0$
($\sigma^{0}$ and $s^{0}$). 
Within on-site interaction, the pair potential  $\Delta_{\rm pair}$ must
fulfill the relation $\Delta_{\rm pair}^{\rm T} = 
- \Delta_{\rm pair}$ owing to the fermionic property. 
We have six possible gap functions classified by a Lorentz-transformation property\cite{NagaiThermal}; 
they are classified into a pseudo-scalar, a scalar, and a polar vector (four-vector). 
In this paper, we focus on a polar vector parallel to $y$-axis
(so-called $\Delta_{4}$\cite{PhysRevLett.105.097001}) given as
\begin{align}
\Delta_{4} &=  \Delta s^{0} \otimes \sigma^{2}, 
\label{eq:deltapair}
\end{align}
motivated by a scenario for explaining the in-plane anisotropy in the
Knight-shift measurement\cite{Hashimoto}. 
The excitation spectrum of this gap function has a point node on
$k_{y}$-axis in the momentum space\cite{NagaiThermal,Nagaivortex,note}.

Now, we propose an intuitive way of understanding the impurity effects
of a point-node gap function. 
Let us exchange the role of spin with the one of orbital, and vice
versa, in the BdG Hamiltonian ($s^{\mu} \leftrightarrow \sigma^{\mu}$,
with $\mu=0,\,1,\,2,\,3$). 
In the dual space, the ``spin'' Pauli (and identity) matrices are written by
$\tilde{s}^{\mu}$, where $\tilde{s}^{\mu}=\sigma^{\mu}$. 
Similarly, the ``orbital'' matrices are denoted by
$\tilde{\sigma}^{\mu}$. 
In the dual space, the topological superconductor with point-nodes, 
$\Delta_{4}$ is regarded as the intra-orbital spin-singlet $s$-wave
pairing, 
\begin{align}
\Delta_{4}^{\rm dual} 
&=  -i \Delta \tilde{\sigma}^{0} \otimes i \tilde{s}^{2}. 
\label{eq:deltapairdual}
\end{align}
The mass term $h_{z}(\bm{k})$ induces the orbital imbalance into the
system in the original space. 
In the dual space, this term is regarded as a contribution inducing the
spin imbalance into the system, 
$h_{z}^{\rm dual} = M(\Vec{k}) \tilde{\sigma}^{0} \otimes
\tilde{s}^{3}$. 
Summarizing the above arguments, we find that in the dual space the
system has a spin-singlet state under the Zeeman magnetic field. 
Under the Zeeman magnetic field, the $s$-wave superconductors become
fragile against non-magnetic impurities, since the spin imbalance due to
the Zeeman magnetic field assists impurities with breaking Cooper
pairs. 
However, when the Zeeman magnetic field is small, the $s$-wave state is
robust against non-magnetic impurities, owing to Anderson's theorem, since 
the non-magnetic impurity is non-magnetic in the dual space.  
Therefore, we claim that the point-node state is not fragile against
non-magnetic impurities in the weak Zeeman orbital field. 

Before checking our statement with a more concrete way, we quantify the
strength of the Zeeman orbital field suitable for studying the role in
the impurity effects. 
For this purpose, we use $\beta$ defined by  
\begin{align}
\beta &\equiv \frac{
|\Vec{P}(\Vec{k}_{\rm F})|}{|M(\Vec{k}_{\rm F})|}, \label{eq:betap}
\end{align}
with $\Vec{P} \equiv (P_{1},P_{2},P_{3})$ and the Fermi wavelength
$\Vec{k}_{\rm F}$. 
The denominator characterizes the Zeeman orbital field, whereas the
numerator is related to the spin-orbit interaction. 
In the dual space, the spin-orbit interaction term $h_{so}(\Vec{k})$ is regarded as the inter-orbital in-plane anisotropic spin-orbit interaction term ($h_{so}^{\rm dual} = \sum_{i=1}^{3} P_{i}(\Vec{k}) \tilde{\sigma}^{i} \otimes \tilde{s}^{1}$). 
With increasing $\beta$, the in-plane spin-orbital interaction
$h_{so}^{\rm dual}$ prevents the spin-polarization along $z$ axis
associated with the dual-space Zeeman magnetic field 
$h^{\rm dual}_{z}(\Vec{k})$. 
It should be noted that $\beta$ is regarded as the indicator of the relativistic effects\cite{NagaiPRBimp}. 
In the non-relativistic region $(\beta \rightarrow 0)$, the effective gap of the nodal topological superconductor is 
the spin-triplet $p$-wave gap with the $\Vec{d}$ vector $\Vec{d} = (v_{z},0,-v_{x})$\cite{NagaiQuasi}. 
Thus, the nodal superconductor is fragile against non-magnetic impurity in the small $\beta$.

Now, let us confirm the robustness numerically with the use of the self-consistent $T$-matrix approximation for impurities\cite{Mahan,Preosti,Balatsky,note2}.  
By considering the randomly distributed non-magnetic impurity potentials [e.g., $V(\Vec{r}) = \sum_{i} \delta(\Vec{r} - \Vec{r}_{i}) V$], 
the $T$-matrix is given as $
T(\Omega) = \left[\openone_{8} - \frac{V}{N}\sum_{\Vec{k}}G_{\Vec{k}}(\Omega) \right]^{-1} V$
with $V = V_{0} \: {\rm diag}\: (1,1,1,1,-1,-1,-1,-1)$, where 
$N$
 is the number of meshes in momentum space. 
The Green's function is 
\begin{align}
G_{\Vec{k}}(\Omega) &= (\Omega-{\cal H}(\Vec{k})- \Sigma(\Omega))^{-1} \equiv
\left(\begin{array}{cc}g_{\Vec{k}}(\Omega) & f_{\Vec{k}}(\Omega) \\
\bar{f}_{\Vec{k}}(\Omega)  & \bar{g}_{\Vec{k}}(\Omega)
\end{array}\right), \label{eq:green}
\end{align}
with the self-energy $
\Sigma(\Omega) = n_{\rm imp} T(\Omega) - n_{\rm imp} V$. 
Here, $n_{\rm imp}$ denotes the impurity concentration. 
We study the impurity effects, checking in-gap states at the low-energy (less than gap amplitude) region in the DOS. 
By solving Eq.(\ref{eq:green}) self-consistently, we obtain the DOS as 
$N(E) = -\frac{1}{2 \pi N} \sum_{\Vec{k}} {\rm tr}\: \left[
{\rm Im} \: \lim_{\eta \rightarrow 0+} g_{\Vec{k}} (E + i \eta)
\right].$
Similarly, we obtain the DOS in the normal states $N_{\rm normal}(E)$, setting $\Delta = 0$. 

\begin{figure}[t]
\vspace{12mm}
\begin{center}
\begin{tabular}{p{ 0.65 \columnwidth}} 
\resizebox{ 0.65 \columnwidth}{!}{\includegraphics{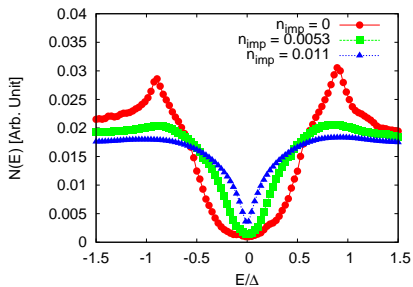}} 
\end{tabular}
\end{center}
\caption{\label{fig:fig1} 
(Color online) Energy dependence of the density of states $N(E)$ in the topological 
gap function with point-nodes, with different impurity concentrations, in strong Zeeman orbital fields ($\beta \sim 0.88$). 
} 
\end{figure}

\begin{figure}[t]
\vspace{10mm}
\begin{center}
\begin{tabular}{p{ 0.65 \columnwidth}} 
\resizebox{ 0.65 \columnwidth}{!}{\includegraphics{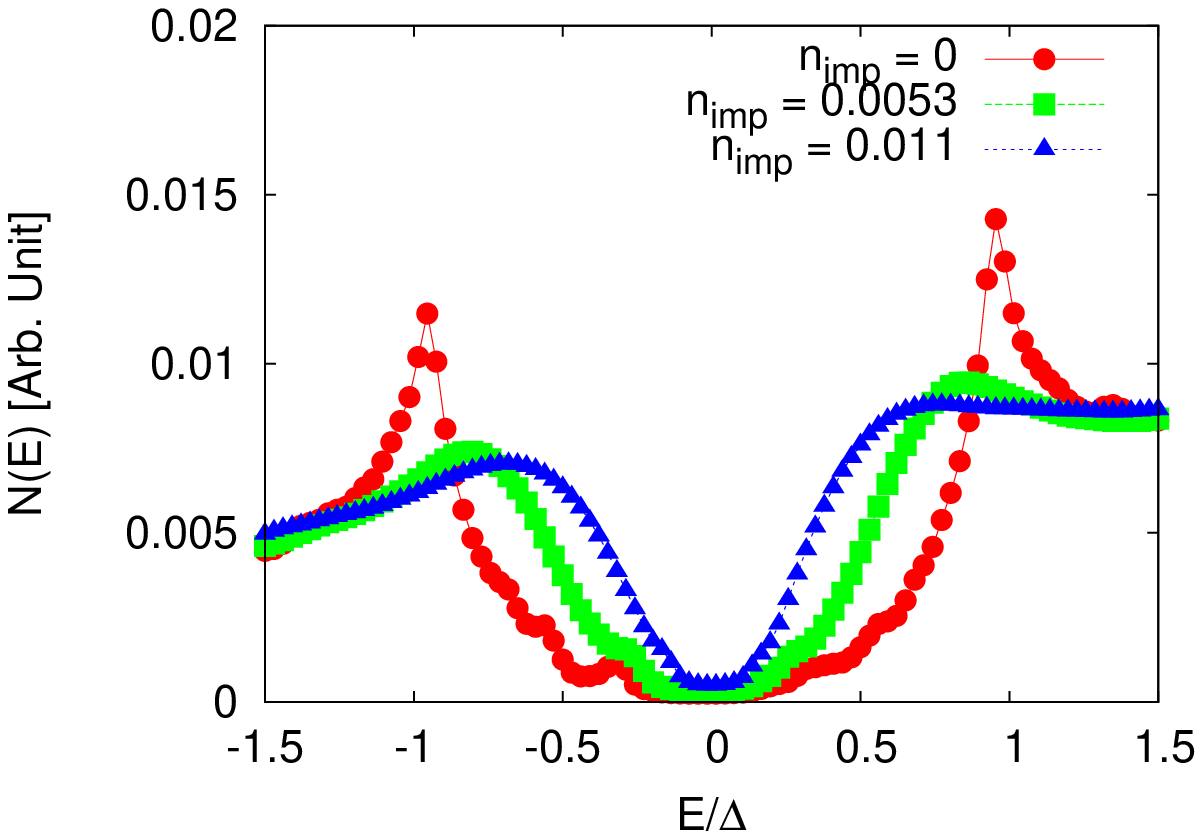}} 
\end{tabular}
\end{center}
\caption{\label{fig:fig2} 
(Color online) Energy dependence of the density of states $N(E)$ in the topological 
gap function with point-nodes, with different impurity concentrations, in weak Zeeman orbital fields ($\beta \sim 2.83$). 
} 
\end{figure}

%
\begin{figure*}[Ht]
\vspace{10mm}
\begin{center}
\begin{tabular}{p{0.77 \columnwidth} p{ 0.77 \columnwidth} } 
\resizebox{0.77  \columnwidth}{!}{\includegraphics{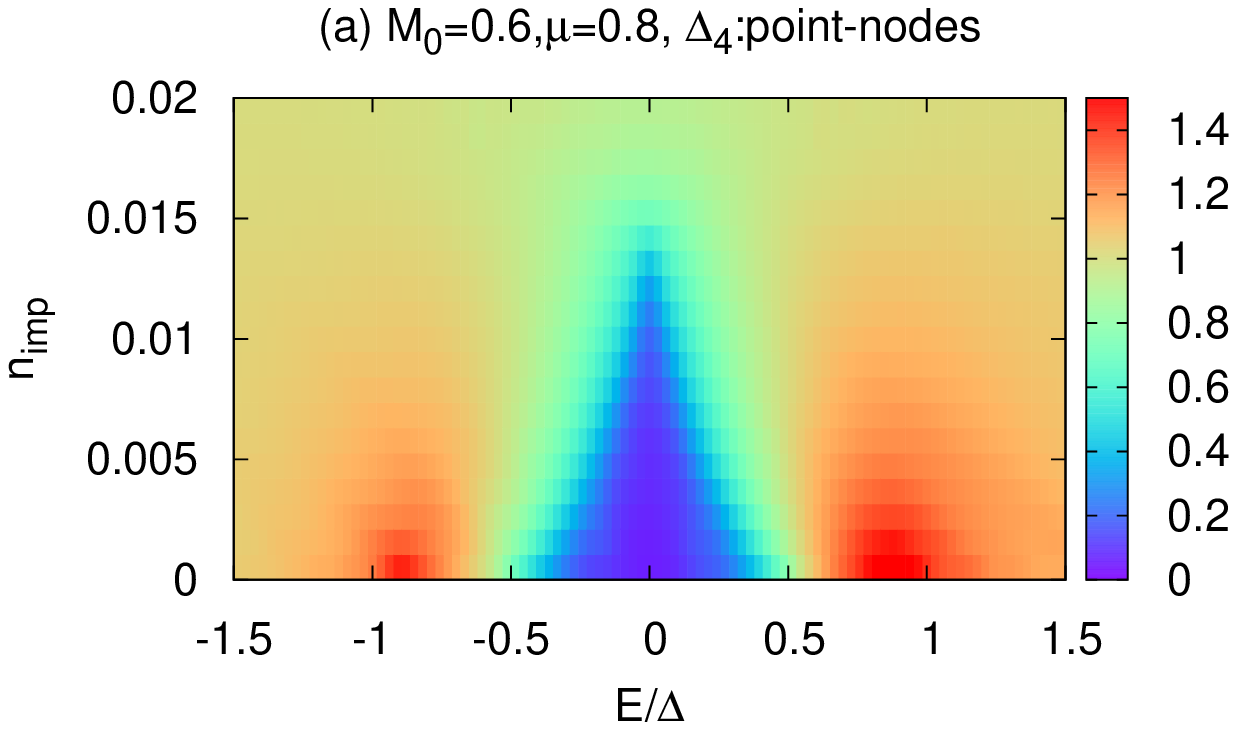}} &\resizebox{ 0.77 \columnwidth}{!}{\includegraphics{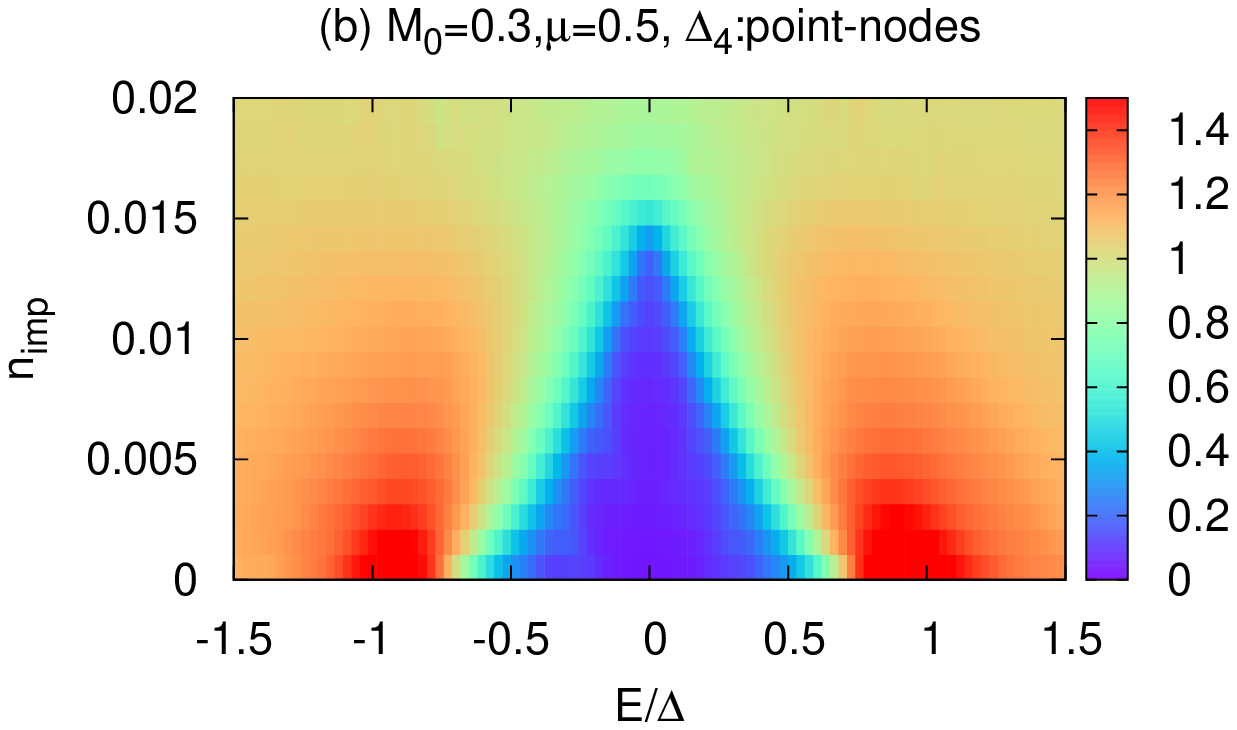}}  \\
\resizebox{0.77  \columnwidth}{!}{\includegraphics{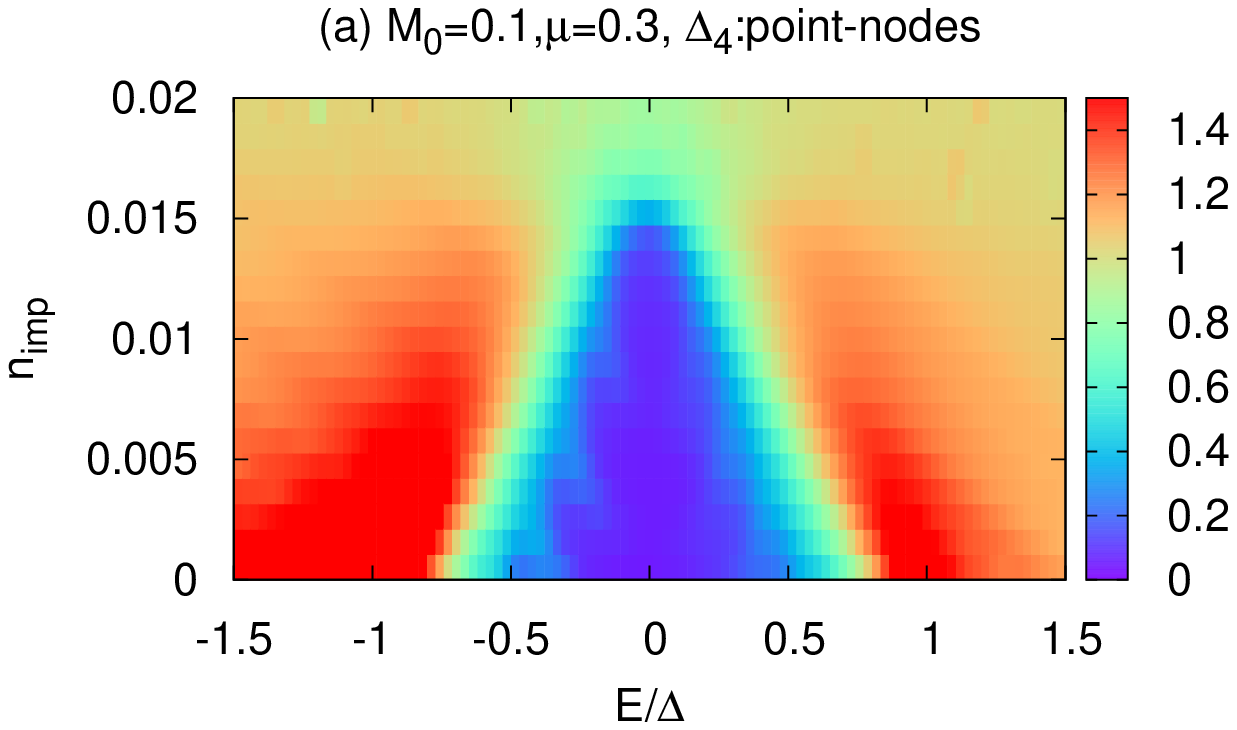}} &\resizebox{ 0.77 \columnwidth}{!}{\includegraphics{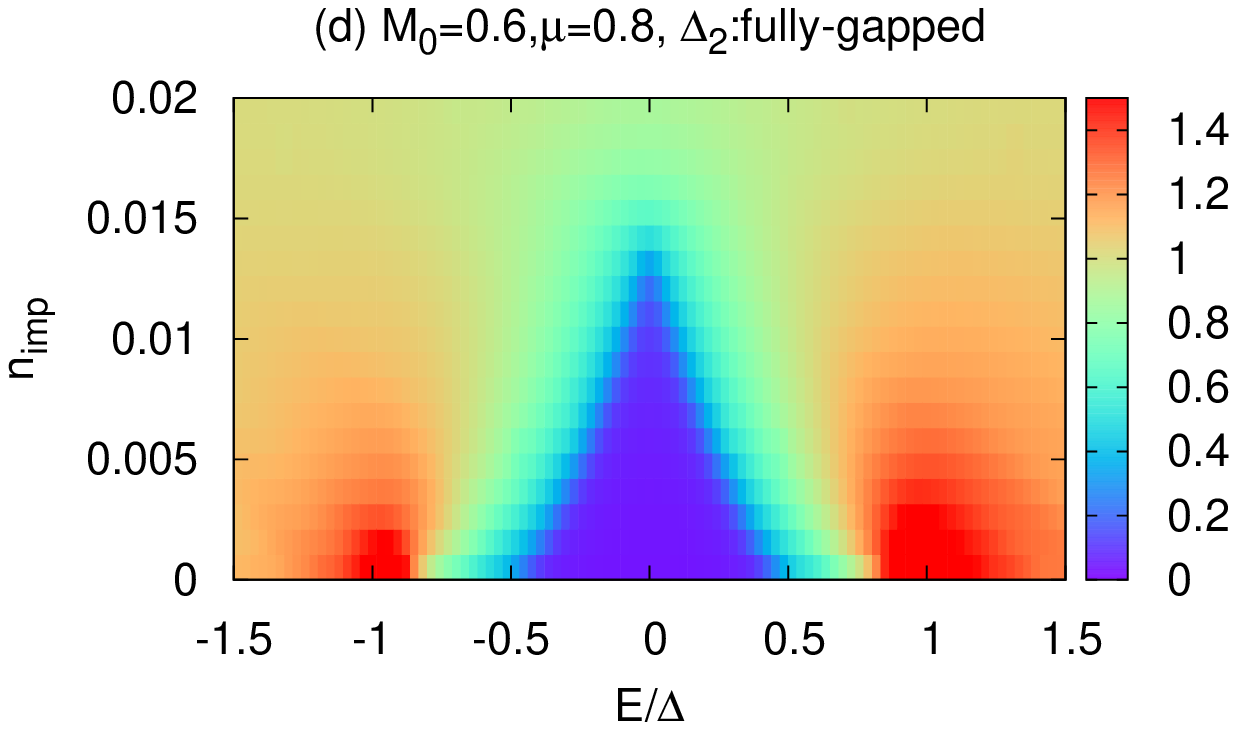}} 
\end{tabular}
\end{center}
\caption{\label{fig:fig3} 
(Color online) 
Non-magnetic impurity-concentration dependence of the ratio of the density of the states (DOS) in superconducting states to that in normal states. 
(a)-(c): The topological superconductors with point-nodes (so-called $\Delta_{4}$) are considered with the different ``indicator'' $\beta = \sqrt{(\mu/M_{0})^{2}-1}$. 
(d):The fully-gapped topological superconductivity (so-called $\Delta_{2}$) is considered. 
The horizontal axis is energy $E/\Delta$ and the vertical axis is the impurity concentration $n_{\rm imp}$. 
The unitary-like scatterer $V_{0} = 10$eV is adopted.  
}
\end{figure*}

\begin{figure}[t]
\vspace{10mm}
\begin{center}
\begin{tabular}{p{ 0.7 \columnwidth}} 
\resizebox{ 0.7 \columnwidth}{!}{\includegraphics{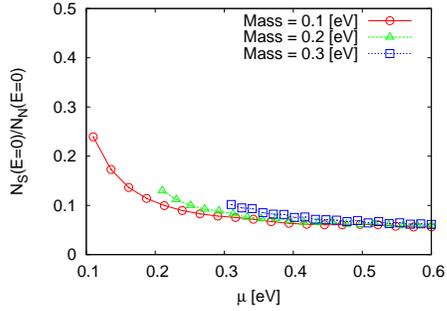}} 
\end{tabular}
\end{center}
\caption{\label{fig:fig4} 
(Color online) The chemical-potential dependence of the ratio of the zero-energy density of the states (DOS) in superconducting states to that in normal states, with the use of the linearized Dirac BdG Hamiltonian. The number of the meshes is $384^{3}$. The impurity concentration is $n_{\rm imp} = 0.02$.
} 
\end{figure}

Let us show the setup for our numerical calculations. 
We adopt the large $\Vec{k}$-mesh size $N = 512^{3}$
in order to accurately describe point-nodes in momentum space.  
We focus on a unitary-like scattering model with $V_{0} = 10$eV, to
study a case that the superconducting pair is broken drastically.  
The gap amplitude is $\Delta = 0.1$eV and the smearing factor is $\eta =
0.0025$eV.  
The unit of energy is eV throughout this paper, unless otherwise noted. 
We set several material variables in the normal-state Hamiltonian
$h_{0}$, using the data from the first-principle calculations of
$\mbox{Bi}_{2}\mbox{Se}_{3}$. 
Typically, the momentum dependence of the coefficients in $h_{0}$ is
described by 
\(
\epsilon(\Vec{k})
=
-\mu
+
\bar{D}_{1} \epsilon_{c}(\veck)
+
(4/3)\bar{D}_{2} \epsilon_{\bot}(\veck)
\) 
and 
\(
M(\Vec{k})
=
M_{0}
- \bar{B}_{1}\epsilon_{c}(\veck)
- 
(4/3) \bar{B}_{2} \epsilon_{\bot}(\veck)
\), with 
\(
\epsilon_{c} = 2 -2 \cos(k_{z})
\), 
\(
\epsilon_{\bot} 
= 3 - 2\cos(\sqrt{3}k_{x}/2) \cos(k_{y}/2) - \cos(k_{y})
\),  
\(
P_{1}(\Vec{k})
=
(2/3)\bar{A}_{2} 
\sqrt{3} 
\sin ( \sqrt{3}k_{x}/2 )
\cos ( k_{y} / 2 )
\), 
\(
P_{2}(\Vec{k})
=
(2/3)\bar{A}_{2} [
\cos ( \sqrt{3}k_{x} /2 )
\sin ( k_{y} /2 ) 
+
\sin ( k_{y} ) 
]
\), 
and 
\(
P_{3}(\Vec{k})
=
\bar{A}_{1} \sin(k_{z})
\). 
The material variables $\bar{D}_{1}$, $\bar{D}_{2}$,
$\bar{B}_{1}$, $\bar{B}_{2}$, $\bar{A}_{2}$, $\bar{A}_{1}$ are
determined by the data from the first-principle calculations in
Ref.~\onlinecite{Zhang:Zhang:2009}. 
The remaining two quantities $M_{0}$ and $\mu$ are variable parameters
in this paper, and they are closely related to the (dimensionless)
strength of the Zeeman orbital field, $\beta^{-1}$.  
For simplicity, we linearize the parameter $\beta$ as $\beta \equiv
A_{2} k_{\rm F}/M_{0} = \sqrt{(\mu/M_{0})^{2}-1}$.  
It should be noted that $\beta$ is equivalent to the indicator of
``relativistic'' effects as shown in our previous
paper\cite{NagaiPRBimp}.  
The Dirac Hamiltonian has two distinct behaviors, depending on $\beta$:
a nonrelativistic limit ($\beta \rightarrow 0$) and an ultrarelativistic
limit ($\beta \rightarrow \infty$).  
We remark that the fully-gapped topological superconductivity (so-called $\Delta_{2}$\cite{PhysRevLett.105.097001}) 
has two aspects, $p$-wave character in a nonrelativistic limit and $s$-wave one in an ultrarelativistic limit, in terms of the robustness against non-magnetic impurities.

First, we show the DOS $N(E)$ with different impurity concentrations. 
We use the mass and chemical potential as $(M_{0},\mu) = (0.6,0.8)$ ({\it i.e.}, $\beta \sim 0.88$), which 
is in strong Zeeman orbital field (or a nonrelativistic region). 
The energy dependence of the DOS without impurities has the power-law like behavior around the zero energy as shown in Fig.~\ref{fig:fig1}. 
On the other hand, 1\% non-magnetic impurities induce in-gap states. 
This result shows that the topological superconductor with point-nodes is fragile against non-magnetic impurities in 
strong Zeeman orbital fields, which is similar to the topological fully-gapped superconductor\cite{NagaiPRBimp,ImpFu}. 
In weak Zeeman orbital fields (a relativistic region) as shown in Fig.~\ref{fig:fig2}, this superconductivity does not have the zero-energy states even with 1\% impurities with the unitary-like scatters whose intensity is one-hundred times larger than the gap amplitude. 
Second, we calculate the impurity-concentration dependence of the ratio of the DOS in 
superconducting states to that in normal states. 
The finite DOS in the low energy indicates the Cooper pairs are broken. 
With increasing the indicator $\beta$ (decreasing the Zeeman orbital field), the blue-colored region becomes large; 
The topological superconductors become more robust, as shown in Fig.~\ref{fig:fig3}. 
We should note that the robustness in the fully-gapped topological superconductor and the nodal topological superconductor is similar to each other. 
Third, we discuss the multidirectional analysis in order to confirm our claim. 
It should be noted that the model Hamiltonian for CuBi$_{2}$Se$_{3}$ is not exactly the linearized Dirac Hamiltonian because of the existence of the higher orders of the momentum. 
Our claim is valid when the material can be described by the linearized massive Dirac Hamiltonian.
The higher orders such as the square of the momentum might cause the different impurity effect we do not consider.  
In addition, it is known that the Fermi-surface evolution from the spheroidal to cylindrical shape appear with 
increasing the chemical potential $\mu$\cite{Mizushima}, which can not appear in the linearized Dirac Hamiltonian.
Thus, with the use of the linearized Dirac BdG Hamiltonian around $\Gamma$ point, we discuss the mass dependence and the chemical-potential dependence to confirm our claim. 
The zero-energy density of states (ZEDOS) with the fixed mass monotonically decreases with increasing the chemical potential as shown in Fig.~\ref{fig:fig4}. 
One can also find that the ZEDOS with the fixed chemical potential monotonically decreases with decreasing the mass.  
These two behaviors consistent with that the indicator $\beta = \sqrt{(\mu/M_{0})^{2}-1}$ characterizes the impurity effects. 
We should note that the values of the ZEDOS with the fixed $\beta$ depend on the mass when the $\beta$ is small. 
This dependence originates from the robustness in the $\Vec{p}$-wave superconductors with the different mass.
The kinetic term in the nonrelativistic limit depends on the mass and $\Delta^{2}/M_{0}$ can not be neglected, which will be discussed in the future. 
These results show that the indicator $\beta$, of the relativistic effects or the Zeeman orbital fields, can well characterize the non-magnetic impurity effects.

Let us discuss the importance of the Zeeman orbital field in terms of the violation of Anderson's theorem. 
Anderson's theorem breaks down when the $\Vec{k}$-averaged anomalous self-energy vanishes\cite{Kopnin:2001} (e.g.,~ in $d$-wave and chiral $p$-wave superconductors). 
The anomalous self-energy with the non-self-consistent Born approximation is 
$\Sigma_{\rm Born}^{\rm A} (\Omega) = -n_{\rm imp} V_{0}^{2}/N \sum_{\Vec{k}} f_{\Vec{k}}(\Omega)$.
In our model, we obtain 
\begin{align}
 \Sigma_{\rm Born}^{\rm A} (\Omega) &=  \frac{n_{\rm imp} \Delta_{4} V_{0}^{2}}{N} \sum_{\Vec{k}} \frac{
 C(\Vec{k})- (M(\Vec{k})+ \Omega)^{2}}
 {D(\Vec{k})}, 
\end{align}
where $C(\Vec{k}) = \Delta^{2} + \epsilon(\Vec{k})^{2}+P_{1}(\Vec{k})^{2}- P_{2}(\Vec{k})^{2}+P_{3}(\Vec{k})^{2}$
 and 
$D(\Vec{k}) = {\rm det} (\Omega - {\cal H}(\Vec{k}))$.
We note that $C(\Vec{k})$ is positive and $D(\Vec{k})$ is strictly negative when $\Omega$ is the Matsubara frequency ($\Omega = i \omega_{n}$), since the $\Vec{k}$-sums of $P_{1}^{2}$ and $P_{2}^{2}$ are same in the normal states and the spectrum of the BdG Hamiltonian is constructed by pairs of positive and negative eigenvalues, owing to its particle-hole symmetry. 
In zero Zeeman orbital fields (i.e. $|M(\Vec{k})| = 0$), the anomalous self-energy never vanishes. 
This anomalous self-energy is very similar to that in the two-dimensional $s$-wave superconductor with the Zeeman magnetic fields, by replacing the Zeeman orbital field  with the Zeeman magnetic field shown in Eq.~(9) in Ref.~\onlinecite{Nagai2Dimp}. 
In strong Zeeman orbital fields ($\beta$ in Eq.~(\ref{eq:betap}) is small), 
$\Sigma_{\rm Born}^{\rm A}$ can be so small that the robustness  dies out. 
Hence, the Anderson's theorem is violated, when the Zeeman orbital field is large.

In conclusion, we studied the robustness against non-magnetic impurities in the topological superconductor with point-nodes, focusing on an effective model of Cu$_{x}$Bi$_{2}$Se$_{3}$. 
We found that the strength of the Zeeman ``orbital'' field (i.e. the indicator $\beta$) characterizes the robustness, since the 
topological superconductor with point-nodes can be regarded as the intra-orbital spin-singlet $s$-wave pairing in the dual space. 
This strength corresponds to the weight of the relativistic effects. 
We showed that the topological superconductivity is not fragile in dirty materials, even with nodes.  
The topological superconductors can not be simply regarded as one
of the {\it conventional} unconventional superconductors.

We thank Y. Ota and M. Machida for helpful discussions and comments. 
The calculations have been performed using the supercomputing 
system PRIMERGY BX900 at the Japan Atomic Energy Agency. 
This study was supported by JSPS KAKENHI Grant Number 26800197.




\begin{thebibliography}{99}
\bibitem{note}
It should be noted that the gap function with point-nodes in in-plane $\theta_{N}$-direction is described as 
$\Delta(\theta_{N}) = \Delta (\sin \theta_{N} s^{0} + i \cos \theta_{N} s^{3}) \otimes \sigma^{2}$.
\bibitem{note2}
We neglect the inter-orbital and spin-orbit scatterings, since the impurities are usually the doped copper atoms in real materials.
\end{thebibliography}
\end{document}